\documentclass[prb,aps,superscriptaddress,twocolumn]{revtex4-2}
\usepackage{graphics,graphicx, array, afterpage}
\usepackage{qcircuit,amsmath}
\usepackage{grffile}
\usepackage[export]{adjustbox}
\usepackage{lineno}
\usepackage{xcolor}
\usepackage{longtable}
\usepackage{braket}
\usepackage{array}
\usepackage{multirow}
\usepackage{enumerate}
\usepackage{amsmath}
\usepackage{amssymb}
\usepackage{esint}
\usepackage{amsthm}
\usepackage{times}
\usepackage{multirow}
\usepackage{comment}
\usepackage{bm}
\usepackage[normalem]{ulem}
\usepackage[colorlinks, citecolor=blue]{hyperref}
\usepackage{tikz}
\usetikzlibrary{arrows.meta}
\newcommand{\ccwexchange}{%
  \mathrel{\vcenter{\hbox{%
    \begin{tikzpicture}[baseline=-0.5ex,>=Stealth]
      \draw[->] (0.55,0.28) to[out=160,in=20] (0,0.28);
      \draw[->] (0,-0.28) to[out=-20,in=-160] (0.55,-0.28);
    \end{tikzpicture}%
  }}}%
}

\hypersetup{linkcolor=magenta,urlcolor=blue,citecolor=blue,pdfstartview={FitH},urlcolor=blue}

\newcommand{\br}{\bm{r}}
\newcommand{\bk}{\bm{k}}

\newcommand{\ba}{\bm{a}}
\newcommand{\bp}{\bm{p}}

\begin{document}

\title{Hidden weak-pairing superconductivity of non-interacting anyons obeying 1/3 statistics}

\author{Zheng-Duo Fan}
\affiliation{Department of Physics, Harvard University, Cambridge, Massachusetts 02138, USA}

\author{Ashvin Vishwanath}
\email{avishwanath@g.harvard.edu}
\affiliation{Department of Physics, Harvard University, Cambridge, Massachusetts 02138, USA}

\author{Zijian Wang}
\email{wzj288@gmail.com}
\affiliation{Institute for Advanced Study, Tsinghua University, Beijing, 100084, China}

\begin{abstract}
We propose that a non-interacting gas of charge-$e/3$ anyons with exchange statistics $\theta=-\pi/3$ can superconduct through a hidden weak-pairing mechanism. Such an anyon gas arises naturally in doped fractional Chern insulators (FCIs) at filling $1/3$ or $2/3$, where projective lattice translations enforce three degenerate anyon pockets. Exploiting this three-pocket structure, we develop a flux-attachment construction in which the average statistical flux vanishes, thereby mapping the problem to three species of composite fermions (CFs) in zero effective magnetic field.  We show that the kinetic energy and statistics of anyons supply the pairing glue and drive the CFs into a $p-\mathrm{i}p$ paired state. The CF strong-pairing phase is adiabatically connected to Laughlin's picture of anyon superconductivity. By contrast, the weak-pairing phase realizes a distinct superconducting phase – its edge is characterized by a chiral central charge $c_-=-1/2$, in contrast to the prediction of an integer $c_-$ based on Laughlin's picture, thereby resolving the discrepancy between previous theories and recent numerics. Our theory provides a natural framework for understanding superconductivity near FCIs, as observed in recent experiments. Finally, we discuss extensions of our theory that predict new chiral superconductors adjacent to FCIs at other fillings.
\end{abstract}

\maketitle
\section{Introduction}
 In quantum mechanics, identical particles confined in two dimensions are not restricted to being either bosons or fermions: their exchange can produce an arbitrary phase $\theta$, defining a class of fractional-statistics particles known as anyons~\cite{wilczek1982quantum}. Seminal early work showed that 
non-interacting anyons with $\theta = \pi(1 - 1/n)$ statistics and (fractional) electric charge have a superconducting ground state~\cite{laughlin1988sc,fetter1989rpa,chen1989anyonsc,lee-fisher1989}. Although anyons are not elementary particles, this theory is experimentally relevant because anyons can emerge as collective excitations of many-body electron systems~\cite{halperin1984statistics,arovas1984statistics,shi2025doping,divic2025anyonsc,han2025anyonsf}. A particularly promising candidate that supports itinerant anyons is fractional Chern insulators (FCIs)~\cite{regnault2011fractional,Cai2023fci,park2023fci,xu2023fci,li2026anyontrion}. For the most prominent $2/3$ filling FCI observed in experiments, the elementary anyon has charge $e/3$ and obeys $\theta =-\pi/3$ (referred to as the $1/3$ anyon for convenience). By pairing two $1/3$ anyons into a $2/3$ anyon ($\theta=2\pi/3$), Laughlin's anyon superconductor theory ($n=3$) becomes applicable, and has already been invoked to explain the superconductivity (SC) recently observed in twisted MoTe$_2$~\cite{xu2025signatures,nosov2026anyonsc,shi2025anyon,pichler2026microscopic}.

However, previous theoretical work leaves an open question and overlooks another fundamental possibility. The open question is what provides the pairing glue for $1/3$ anyons \cite{jain2025molecular,xu2025anyoncluster,khalaf2026bound,taige2026anyonmolecule,pichler2026microscopic}. The overlooked question arises upon careful re-examination of the concept of pairing in BCS theory \cite{bcs1957}: When two fermions pair into a boson, it is important to distinguish two pairing pictures — pairing in momentum space (where momentum $\bk$ pairs with $-\bk$) and pairing in real space (a dilute gas of tightly bound pairs). Although these two pictures are sometimes adiabatically connected (the BCS--BEC crossover)~\cite{eagles1969possible,leggett1980modern}, they describe genuinely distinct phases whenever the momentum-space pairing is topologically non-trivial (the pair wave function winds in momentum space), as has been thoroughly studied in the context of topological superconductors~\cite{read2000paired,Alicea:2012review}. The momentum-space and real-space pairings are referred to as weak and strong pairing, respectively. Returning now to our anyon superconductor problem: at first glance, only the real-space picture seems available, since anyons do not possess a Fermi surface, and indeed previous theoretical work has assumed that $1/3$ anyons are simply tightly bound together in real space. But this needs further examination. Could a weak-pairing phase – a new anyon superconducting phase – have been hiding all along, obscured by the lack of an obvious Fermi surface?

In this work, we address both of the above questions.  We begin with a flux-attachment construction that utilizes the existence of three distinguishable anyon pockets in doped FCIs.  We attach pocket-dependent statistical flux to the composite fermions (CFs), in such a way that the average statistical flux vanishes.  This maps the $1/3$ anyon gas to three CF pockets living in zero field. With this construction in place, the answer is nearly immediate. The statistical gauge field fluctuation itself provides a natural pairing glue for the $1/3$ anyons, without requiring an additional pairing interaction.  Since the CFs in zero field have Fermi surfaces, both strong and weak pairing become accessible: the strong-pairing phase is adiabatically connected to the conventional $2/3$-anyon superconductor, whereas the weak-pairing phase realizes a distinct superconducting phase. We show that this newly identified phase is a charge-$2e$ $f-\mathrm{i}f$ superconductor with chiral central charge $c_-=-1/2$. We further apply our theory to the superconducting phase recently observed in twisted MoTe$_2$. Finally, we predict new chiral superconductors in doped $\nu=1-1/m$ or $\nu=1/m$  ($m=5,\,7,\cdots$) FCIs beyond $2/3$ filling. Compared to previous theories~\cite{shi2025doping,shi2025anyon,shi2025nonabelian,nosov2026anyonsc,pichler2026microscopic,Ma_2020,zhang2026holon,lotric2026}, our theory has the following advantages: 1. We provide a natural mechanism for pairing/superconductivity---the kinetic energy and statistics of anyons provide the pairing glue. 2. The weak-pairing phase exposed under our flux attachment construction agrees with recent numerical results on superconductivity near FCIs, where the same pairing symmetry and chiral central charge were identified~\cite{guerci2026topological,taige2025chiral}, in contrast to the $c_-=-2$ prediction of the conventional theory of anyon superconductivity~\cite{shi2025doping,nosov2026anyonsc}. 3. Our construction utilizes minimal flux attachment, thereby reducing gauge fluctuations and enhancing the stability of the mean field ansatz.

\section{Flux-attachment Construction}
We first recall Laughlin's construction for $2/3$ anyon gas~\cite{laughlin1988sc}: Each $2/3$ anyon is represented as a fermion (the CF) carrying statistical gauge flux $\Phi=-2\pi/3$, and thus the problem is reformulated as CFs coupled to statistical gauge field. Then applying mean-field approximation by smearing the flux is smeared to its average value, one gets CFs in a uniform (fictitious) magnetic field, which form a $\nu =-3$ integer quantum Hall states. Finally, to translate the CF resistivity to the physical resistivity, one needs to add an additional Hall response term inherited from the flux attachment: a nonzero current of CFs leads to a flux current, which induces a transverse electric field (of the statistical gauge field)~\cite{zhang1992cslg2}. This additional Hall response exactly cancels the Hall resistivity of CFs, and leads to zero physical resistivity. For $1/3$ anyon gas, however, the recipe breaks down: it leads to partially filled CF Landau levels, leaving no simple mean-field ansatz. In the following, we propose a new flux attachment, where the average statistical flux vanishes. 

\begin{figure}[htb]
\centering
\includegraphics[width=1.0\linewidth]{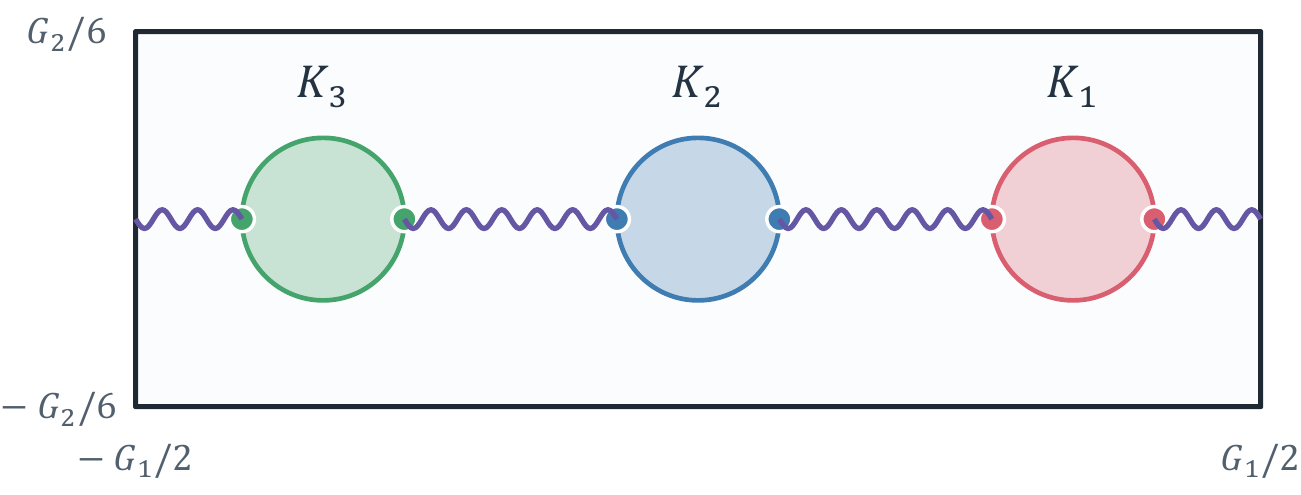}
\caption{\textbf{Schematic of anyon pockets in the reduced Brillouin zone.} The three pockets centered at $K_1, K_2, K_3$ are shown in red, blue and green, respectively. Purple wavy lines illustrate inter-pocket pairing on the Fermi surfaces.} 
\label{fig1}
\end{figure}

Our construction is made possible by the fact that $1/3$ anyons in doped FCIs (for concreteness we focus on hole filling $\nu =2/3+\delta$, aligned with the setup in recent experiments~\cite{xu2025signatures}) carry a projective representation of lattice translation symmetry: $T_1T_2=e^{-2\pi i/3}T_2T_1 $, where $T_\alpha$ translates by the primitive lattice vector ${\bm l}_\alpha$~\cite{cheng2016translational}. As a result, the anyon band is labeled by eigenvalues of $T_1, T_2^3$, and its dispersion has three-fold degeneracy in the reduced Brillouin zone~\cite{gonccalves2026spinless}.
For small doping $\delta\ll 1$, we get three anyon pockets near the valley momenta $\bm{K}_{i=1,2,3}$ related by $T_2$, $\bm{K}_{i+1}=\bm{K}_i-\bm{G}_1/3$, with $i+1$ understood modulo $3$, and $\bm{l}_\alpha\cdot \bm{G}_\beta =2\pi \delta_{\alpha\beta}$. See Fig.~\ref{fig1} for an illustration. We introduce one CF $\psi_{i}$ for each pocket, and assume that the density distribution respects the translation symmetry: $\langle n_i\rangle= {\bar n}=\frac{\delta}{|\bm l_1\times \bm l_2|}~(n_i\equiv \psi_i^\dagger\psi_i)$. 

We now describe our flux attachment, where we use a freedom that is absent in a single-flavor anyon gas. Let $\Phi_{ij}$ be the Aharonov-Bohm phase acquired when a $\psi_i$ particle winds once around a $\psi_j$ particle.  For particles in the same pocket, exchange is meaningful and gives
\begin{equation}
{\Phi_{ii}\over 2}+\pi=-{\pi\over 3}\quad (\mathrm{mod}\;2\pi).
\label{eq:intra_statistics}
\end{equation}
where the $\pi$ comes from the fermion exchange statistics.
On the other hand, particles in different pockets are distinguishable, and only their full mutual braid is physical:
\begin{equation}
\Phi_{i\neq j}=-{2\pi\over 3}\quad(\mathrm{mod}\;2\pi).
\label{eq:inter_statistics}
\end{equation}
The uniform choice $\Phi_{ij}=4\pi/3$ satisfies these conditions, but leads to CFs in partially filled Landau levels at mean-field level. However, by noticing that Eq. \eqref{eq:inter_statistics} only fixes the off-diagonal entries mod $2\pi$, we can shift each $\Phi_{i\neq j}$ by $-2\pi$:
\begin{equation}
\frac{1}{2\pi}\Phi=\begin{pmatrix} \frac{2}{3} & -\frac{1}{3} & -\frac{1}{3} \\ -\frac1{3} & \frac{2}{3} & -\frac{1}{3}\\ -\frac{1}{3} & -\frac{1}{3} &\frac{2}{3}\end{pmatrix}
\label{eq:Phi_ij}
\end{equation}
In this construction, $\sum_{j}\Phi_{ij}=0$, which means that the average flux seen by each CF is zero.

As is standard, flux attachment can be implemented by coupling CFs to Chern-Simons gauge fields~\cite{zhang1989cslg1,zhang1992cslg2,halperin1993half,lopez1991fractional}, with the generic Lagrangian:
\begin{equation}
\begin{gathered}
\mathcal{L}=\sum_i\psi_i^\dagger D_{q_{iI}a_I}\psi_i
-\frac{1}{4\pi}K_{IJ}a_I da_J\\
 D_{q_{iI}a_I}\equiv\Big[\mathrm{i}\partial_t+q_{iI}a_{I0}-\frac{(-\mathrm{i}\nabla-q_{iI}\bm{a}_I)^2}{2m}\Big ]\,,
\end{gathered}
\label{eq:CF_Lagrangian1}
\end{equation}
where $ada\equiv \epsilon^{\mu\nu\rho}a_\mu \partial_\nu a_\rho$. Since the rank of $\Phi_{ij}$ is 2, we need two statistical gauge fields to obtain Eq.~\eqref{eq:Phi_ij} (we use $I,J=1,2$ for the gauge field indices). The $K$ matrix and the gauge charge vectors $q_i$ of our construction are: 
\begin{equation}
K=\begin{pmatrix} 2 & 1 \\ 1 &2\end{pmatrix},\quad q_{1}=\begin{pmatrix} 1\\ 0\end{pmatrix},\, q_2=\begin{pmatrix} 0\\ 1\end{pmatrix},\,  q_3 =\begin{pmatrix} -1\\ -1\end{pmatrix}
\label{eq:K_matrix}
\end{equation}
 This Chern-Simons field theory describes the following flux attachment setting: $\frac{\nabla\times a_1}{2\pi}=\frac{2}{3}n_1-\frac{1}{3}n_2-\frac{1}{3}n_3$, $\frac{\nabla\times a_2}{2\pi}=-\frac{1}{3}n_1+\frac{2}{3}n_2-\frac{1}{3}n_3$. One can verify that this flux attachment construction gives the Aharonov-Bohm phase $\Phi_{ij}=2\pi q^T_iK^{-1}q_j$ as in Eq.~\eqref{eq:Phi_ij}.

To implement the intrinsic hard-core condition on the anyons, one needs to add a contact interaction term
$-\frac{2\pi}{3m}\sum_{i<j}\psi_i^\dagger(x)\psi_j^\dagger(x)\psi_j(x)\psi_i(x)$
to the Lagrangian~\cite{Kim1997contact}.
With this term, the Lagrangian \eqref{eq:CF_Lagrangian1} gives an exact description of the 3-pocket non-interacting anyon gas. In particular, it retains the valley $PSU(3)$ symmetry expected for the non-interacting anyon gas~\cite{shi2026superconductivity}, which acts on both $\psi_i$ and $a_I$.

As in previous approaches~\cite{shi2025doping,shi2025anyon,shi2025nonabelian,nosov2026anyonsc,pichler2026microscopic,Ma_2020,zhang2026holon,lotric2026}, we will adopt a mean-field approximation and treat fluctuations around the mean-field saddle point as weak. Our flux-attachment construction has the following two advantages:

\begin{itemize}
\item  At the mean field level, $\bar{a}_1=\bar{a}_2=0$, so that CFs in each pocket form a Fermi sea, providing a natural mean field saddle to study the pairing instability of CFs.
\item   Our construction realizes minimal flux attachment on each CF. As a result, it leads to minimal gauge fluctuations, enhancing the stability of the mean field saddle point.
\end{itemize}

Our mean field ansatz breaks the $PSU(3)$ symmetry, which if correct suggests that the ground state of non-interacting anyons also spontaneously breaks this symmetry, which we will elaborate on in an upcoming publication.

\section{Pairing by Statistical Gauge fields}

In this section, we start from the 3-pocket CF metal obtained in mean-field approximation, and discuss how gauge fluctuations can lead to pairing between CFs. CF pairing mediated by statistical gauge fields has also been studied in half-filled Landau levels~\cite{greiter1992paired,metlitski2015cooper} and bilayer Quantum Hall systems~\cite{bonesteel1996gauge,fu2017interlayer,halperin2021bilayer}.
It is convenient to work in the second quantized Hamiltonian,
\begin{equation}
\hat H=\sum_i\int\!d^2\br\,\hat \psi^\dagger_i(\br)\frac{1}{2m}\left(\hat{\bp} - q_{iI}\hat\ba_I(\br)\right)^2\hat \psi_i(\br)
\label{eq:H_CF}
\end{equation}
where
\begin{equation}
\hat \ba_I(\br)=(K^{-1})_{IJ}q_{iJ}\int\!d^2\br'\,\frac{\hat{z}\times(\br-\br')}{|\br-\br'|^2} \hat n_i(\br')
\label{eq:gauge_field_explicit}
\end{equation}
Now we substitute the above expression for $\hat\ba_I(\br)$ into the Hamiltonian, and decompose the Hamiltonian into three parts, $\hat H=\hat H_0+\hat H^{(p\cdot a)}+\hat H^{(a^2)}$. Here $\hat H_0=\sum_{i,\bk}\frac{\bk^2}{2m}\hat \psi_i^\dagger(\bk)\hat \psi_i(\bk)$ is the kinetic term. In the following, we show that both $\hat H^{(p\cdot a)}$ and $\hat H^{(a^2)}$ act as pairing interactions. 

We begin with the $\hat H^{(a^2)}$ term. It is a $6$-fermion term, and as in Ref.~\cite{chen1989anyonsc}, we make the approximation $\hat H^{(a^2)}\approx \sum_i \frac{\bar n}{2m} (q_{iI}\hat \ba_I)^2$, which leads to
\begin{equation}
\hat H^{(a^2)}\approx(\frac{1}{3}-\delta_{ij})\frac{\pi\bar{n}}{m}\int\!d^2\br\!\int\!d^2\br' \ln(|\br-\br'|) \hat n_i(\br)\hat n_j(\br')
\end{equation}
This term takes the form of an effective Coulomb interaction: it is repulsive within the same pocket and attractive between different pockets. This sign structure can be understood by the following: Because CFs in different pockets carry different/opposite gauge flux, bringing two CFs in same pocket together reinforces the statistical gauge field, while bringing CFs in two different-pockets together reduces the field strength.  Since the $\hat\ba^2$ term gives an energy density of the gauge field, it leads to intra-pocket repulsion and inter-pocket attraction.

Now we analyze the other term:
\begin{equation}
\hat  H^{(p\cdot a)} = -\frac{\Phi_{ij}}{2\pi}\int\!d^2\br\!\int\!d^2\br'\hat{\bm{j}}_i(\br)\cdot\frac{\hat{z}\times(\br-\br')}{|\br-\br'|^2}\hat n_j(\br')
\end{equation}
where $\hat{\bm{j}}_i(\br) = \hat \psi_i^\dagger(\br)\frac{\hat{\bp}}{m}\hat \psi_i(\br)$. This term is a current-density interaction. It breaks time-reversal symmetry and selects the chirality of pairing. {color{red} Fourier transforming it to momentum space and keeping only the term in the Cooper channel, we obtain
\begin{equation}
\begin{aligned}
\frac{1}{2\mathcal{A}}\sum_{i,j,\bk,\bk'}V^{(p\cdot a)}_{ij;\bk\bk'}&\hat \psi^\dagger_i(\bk)\hat \psi^\dagger_j(-\bk)\hat \psi_j(-\bk')\hat \psi_i(\bk'),   \\ 
V_{ij;\bk\bk'}^{(p\cdot a)} &= -\frac{2\Phi_{ij}}{m}\frac{\mathrm{i}(\bk'\times\bk)\cdot\hat{z}}{|\bk'-\bk|^2}      
\end{aligned}
\end{equation}
where $\mathcal{A}$ is the total area of the system. On the circular Fermi surface, we define $\varphi =  \phi_{\bk}-\phi_{\bk'}$, and then $V^{(p\cdot a)}_{ij;\bk\bk'}= V^{(p\cdot a )}_{ij}(\varphi)=\frac{-\mathrm {i}\Phi_{ij}}{m}\cot \frac{\varphi}{2}$. Its $l$-angular-momentum component is: 
\begin{equation}
V^{(p\cdot a)}_{ij;l}=\mathrm{P.V.}\int_{-\pi}^{\pi} \frac{d\varphi}{2\pi} e^{-il\varphi } V_{ij}^{(p\cdot a)}(\varphi)=-\frac{\Phi_{ij}}{m}\text{sign}(l)
\label{eq:V_l}
\end{equation}
}
Thus it provides intra-pocket (inter-pocket) attraction in the positive (negative) angular-momentum channel.

From the above, we are ready to determine the dominant pairing channel. We write the pairing order parameter as $\langle \hat \psi_i(-\bk) \hat \psi_j(\bk) \rangle = \Delta_{ij}(\bm k)$. 
For inter-pocket pairs, both types of interactions contribute to pairing and cooperate in the negative angular momentum channel. This indicates that inter-pocket pairing is energetically favored.
What about pairing symmetry? It is determined by the following three factors: (i) For even angular momentum, $\Delta_{ij}(\bm k)$ is even in $\bm k$, and Fermi statistics requires $\Delta_{ij}(\bm k)=-\Delta_{ji}(\bm k)$, so $\text{rank}(\Delta)<3$, which means there is at least one unpaired Fermi surface. Therefore, to obtain full pairing, we only consider odd angular momentum. (ii) the density-current interaction selects negative angular momentum pairing. (iii) the density-density interaction – attractive for all momentum transfers – is strongest for lowest possible angular momentum.

Combining the above three factors, we now come to our final conclusion: The statistical gauge field mediates an inter-pocket, $p-\mathrm{i}p$ pairing. The inter-pocket pairing is depicted in Fig.~\ref{fig1}. It is worth emphasizing that this pairing is driven by the kinetic energy and statistics of anyons, without the need for additional interaction (the contact interaction plays no role in any non-zero angular momentum channel). In principle, a subdominant intra-pocket pairing is symmetry allowed and will be induced, but the conclusions will not change as long as the inter-pocket pairing is dominant. More explicitly, we expect the CF pairing order parameter to have the form
\begin{equation}
\begin{aligned}
\langle \hat \psi_i(-\bk) \hat \psi_j(\bk) \rangle &=\big[(1-\delta_{ij})\Delta_{o}(k)+\delta_{ij}\Delta_d(k)  \big]e^{-\mathrm{i}\phi_{\bm k}} \,,\\
|\Delta_{o}(k)|&\gg |\Delta_{d}(k)|
\end{aligned}
\end{equation}
where $k\equiv |\bm k|$ and  $\phi_{\bm k}$ denotes the polar angle of $\bm{k}$.

\section{From CF Pairing to Physical Superconductivity}
In the previous  section we show that the CFs form a superconducting state. In this section, we show what this means for the physical electron system. This is not obvious, since statistical transmutation usually drastically changes the transport properties, e.g., in Laughlin's theory, integer quantum Hall states of CFs lead to anyon superconductors. To examine the physical electromagnetic response, we add coupling to the background electromagnetic field $A$ (taking $e=\hbar =1$):
\begin{equation}
\mathcal{L}=\sum_i\mathcal \psi_i^\dagger D_{q_{iI}a_I+A/3}\psi_i
-\frac{1}{4\pi}K_{IJ}a_I da_J+\frac{2/3}{4\pi}A dA \,.
\label{eq:CF_Lagrangian2}
\end{equation}
The CFs are coupled to $A/3$ since each $1/3$ anyon carries charge $e/3$~\cite{laughlin1983}, and the background $\frac{2/3}{4\pi}AdA$ term comes from the Hall conductance of the parent $\nu =2/3$ FCI. The Dirac quantization condition is $\int_{\Sigma} d(a_I+A/3)\in 2\pi \mathbb{Z} $, for any closed surface $\Sigma$. To restore the usual quantization of $da$, we must shift $a_I\rightarrow a_I+A/3$ to obtain integer levels as in Eq.~\eqref{Eq:Final}.

We denote the particle current carried by $\psi_i$ to be $\bm{J}_i$, with the corresponding single-pocket conductivity $\sigma_{\psi_i}$ given by $\bm{J}_i = \sigma_{\psi_i}\big(q_{iI}\bm{e}_I+e\bm E/3\big)$, where $\bm{e}_I\equiv\nabla a_{I0}-\partial_0 \bm{a}_I$ is the Chern-Simons counterpart of electric field. Due to translation invariance, $\sigma_{\psi_1}=\sigma_{\psi_2}=\sigma_{\psi_3}=\sigma_{\psi}/3$, where $\sigma_{\psi}\equiv \sum_i \sigma_{\psi_i}$ is the total conductivity of CFs. Since each $\psi_i$ carries electric charge $e/3$, and $\sum_i q_i=(0,0)^T$, the total electric current produced by anyons is $\frac{e}{3} \sum_i \bm{J}_i= \frac{e^2}{9}\sigma_\psi \bm{E}$. Combining it with the background Hall response of FCI, we obtain the physical conductivity tensor:
\begin{equation}
\sigma =\frac{2e^2}{3h}\begin{pmatrix} 0 & 1\\ -1 & 0\end{pmatrix} +\frac{e^2}{9}\sigma_\psi \,.
\label{eq:phys_conductivity}
\end{equation}
Therefore, in our construction, superconductivity of CFs leads to a physical superconductor. This is consistent with the intuition that when $1/3$ anyons pair into $2/3$ anyons, the itinerant $2/3$ anyon gas can then form a superconductor. 

Remarkably, we can now conclude that the ground state of non-interacting $1/3$ anyon gas is a superconductor, though it does not belong to the series $\theta =(1-1/n)\pi$ identified by Laughlin. Compared to Laughlin's framework, the only essential additional input we need is the existence of three anyon pockets, which arise automatically when the anyons emerge from an electron system at 2/3 filling.

\section{Chern-Simons-Landau-Ginzburg Effective Action}
In this section, we further study the properties of the superconducting phase through the Chern-Simons-Landau-Ginzburg effective action~\cite{zhang1989cslg1,zhang1992cslg2}. We now specialize to the energetically favorable inter-pocket pairing of CFs. We start with the order parameter of CFs: $\Delta_{ij}=\psi_i\psi_j\equiv |\Delta_{ij}|e^{\mathrm{i}\theta_{ij}}$. Importantly, the CF Cooper pairs carry both electric charge $2e/3$, and Chern-Simons gauge charge $\tilde q_{ij}=q_i+q_j$. For convenience, we introduce the pair index $p\equiv ij =(12),(23),(31)$. When the CFs condense, we obtain the phase-only low-energy action:
\begin{equation}
\begin{aligned}
\mathcal{L}= \sum_p &
\frac{\chi_s}{2}\left[
\left(\partial_0\theta_{p}-\alpha_{p0}\right)^2 -v^2 \left(\nabla \theta_p-\bm \alpha_{p}\right)^2\right]   \\     
&-\frac{1}{4\pi}K_{IJ}a_I da_J+\frac{2/3}{4\pi}AdA
 \,,
 \end{aligned}
\label{eq:CSLG}
\end{equation}
where $\alpha_{p\mu}\equiv \tilde q_{pI}a_{I\mu}+2A_\mu/3$, with $\tilde q_{12}=(1,1)^T$, $\tilde q_{31}=(0,-1)^T$, $\tilde q_{23}=(-1,0)^T$.  The parameters $\chi_s,v$ are non-universal. 

To avoid a vortex with divergent energy cost, for any large circle around vortex core, 
\begin{equation}
\oint (\nabla\theta_{p} -\tilde q_{pI}\bm{a}_I-\frac{2}{3}\bm{A})\cdot \bm{dl} =0\,.
\label{eq:vortex_condition}
\end{equation}
Summing the equations for all three $p$, we get $\oint(\nabla \Theta-2\bm{A})\cdot \bm{dl}=0$, where $\Theta =\sum_p \theta_p$ is the gauge-neutral phase field. The minimal magnetic flux quanta of the superconductor is $\oint \bm{A}\cdot \bm dl=\pi$, i.e., $h/2e$ in physical units. Thus the inter-pocket paired state is a charge-$2e$ superconductor. 

The same observation also identifies the physical SC order parameter. Each individual $\Delta_{ij}$ is not gauge invariant, and cannot serve as local order parameters. Nevertheless, the SC order parameter can be constructed from their combination
\begin{equation}
{\cal O}_{\rm SC}(\bm{R},\bm{r})\sim  e^{2\mathrm{i}\sum_i\bm K_i\cdot \bm R}\Delta_{12}(\bm{R},\bm{r})\Delta_{23}(\bm{R},\bm{r})\Delta_{31}(\bm{R},\bm{r})\,,
\label{eq:SC_order_parameter}
\end{equation}
where $\bm R,\bm r $ denote the center-of-mass and relative coordinate of the Cooper pair, respectively. It is gauge invariant, and carries charge $2e$. 
From Eq.~\eqref{eq:psi_translation}, one can further show that ${\cal O}_{\rm SC}(\bm{R},\bm{r})$ transforms as an ordinary local operator: ${\cal O}_{\rm SC}(\bm{R},\bm{r})\rightarrow {\cal O}_{\rm SC}(\bm{R}+\bm l_\alpha,\bm{r})$ under $T_\alpha$. We can also identify the physical pairing symmetry: Since the CF Cooper pairs each carry angular momentum $l=-1$, the angular momentum of physical Cooper pairs should be $-3$, which indicates $f-\mathrm{i}f$ pairing. Notably, we have invoked the emergent continuous rotation symmetry in the low-energy theory.
In terms of the microscopic hole annihilation operator $\hat c$, we expect the superconducting order parameter $ \langle c(\bm R+\bm r/2)c(\bm R-\bm r/2)\rangle \sim e^{2\mathrm{i}\sum_i\bm K_i\cdot \bm R}|\Delta(\bm{r})|e^{-3\mathrm{i}\phi_{\bm r}}$ at distance $|\bm r|$ much larger than the lattice constant. The state is a uniform SC if $\left(2\sum_i\bm K_i\right)$ is a reciprocal lattice vector, and a pair-density-wave SC otherwise.

Next, we discuss the low-energy excitations. We start with three $U(1)$ phase fields, $\theta_{12},\theta_{31},\theta_{23}$. After coupling to the Chern-Simons fields, the two relative phase fields are Higgsed together with $a_1,a_2$, and only the gauge-invariant total phase $\Theta$ remain ``uneaten". This can be identified with the Goldstone mode of the global $U(1)$ symmetry, bypassing the need for involved RPA calculations in the conventional formulation of anyon SC~\cite{fetter1989rpa,chen1989anyonsc}. 
After integrating out the massive degrees of freedom, we get the low-energy effective theory 
\begin{equation}
\mathcal{L}=\frac{\chi'_s}{2}\left[ (\partial_0\Theta-2A_0)^2-v^2(\nabla\Theta-2\bm A)^2\right]\,.
\end{equation}

Finally, this inter-pocket paired state has no deconfined anyon excitations.  First, the original $1/3$ anyon is confined: each $1/3$ anyon is represented as a CF carrying fractional statistical fluxes specified by Eq.~\eqref{eq:Phi_ij}. Therefore, an isolated $1/3$ anyon would lead to violation of Eq.~\eqref{eq:vortex_condition} and produces a divergent energy cost. Second, no new deconfined anyons emerge from the Chern-Simons gauge fields. The reason is that the inter-pocket CF pairs all carry primitive gauge charges, i.e., they can generate arbitrary integer gauge charges. Hence the condensate completely Higgses the $U(1)^2$ Chern-Simons fields, leaving behind no residual discrete deconfined gauge subgroup. In Appendix \ref{app:pv_duality} we use particle-vortex duality to explicitly show that all dynamical flux excitations obey trivial statistics.

\section{Chiral Central Charge}
The Chern-Simons-Landau-Ginzburg theory above captures long-wavelength phase fluctuations and vortex excitations of the superconducting state. However, it does not distinguish different types of CF pairings. There are two topologically distinct $p-\mathrm{i}p$ superconducting phases for CFs: a strong-pairing phase, where the CF Cooper pair wavefunction $g(\bm{r})\sim e^{-|\bm{r}|/\xi}e^{-\mathrm{i}\phi_{\bm{r}}}$, where $\bm{r}$ is the relative coordinate of the pair, and a weak-pairing phase $g(\bm{r})\sim \frac{e^{-\mathrm{i}\phi_{\bm{r}}}}{|\bm r|} $~\cite{read2000paired}. In the following, we discuss the strong- and weak-pairing phases respectively, and show that they lead to distinct chiral central charges $c_-$ for the physical edge theory, which determine the experimentally measurable thermal Hall conductance $\kappa_{xy}=\frac{\pi}{6\hbar}c_-k_B^2T$~\cite{kane1997thermal,read2000paired,kitaev2006anyons}.

We first show that the strong-pairing phase is adiabatically connected to Laughlin's $2/3$ anyon superconductors. To gain some intuition, we can consider the BEC limit $\xi\ll \bar n^{-1/2}$, where CFs in different pockets form tightly bound bosons. In terms of the original anyons, this is precisely the molecular limit in which two $1/3$ anyons bind into a $2/3$ anyon, and, as demonstrated above, they do form a charge-$2e$ superconductor with no intrinsic deconfined anyon excitations. Next, we calculate the chiral central charge. We first note that at zero doping, the $\nu =2/3$ FCI is non-chiral ($c_-=0$). However, since $K$ has two positive eigenvalues, the Chern-Simons sector would leave two chiral bosons on the edge~\cite{wen1995topological}. Thus, there must be an additional background term to compensate for this. In the strong-pairing phase of CFs, the Chern-Simons fields are Higgsed, and the CFs contribute no chiral edge mode, so we get $c_-=-2$ from the background term. This is also consistent with recent studies of the $2/3$ anyon superconductor~\cite{shi2025doping,nosov2026anyonsc,pichler2026microscopic}.

Then we turn to the weak-pairing phase, which corresponds to the case where pairing only happens near the CF Fermi surface. In this case, the $1/3$ anyons no longer form two-body bound states in real space. Rather, weak pairing is a many-body collective phenomenon that can only occur when anyons are itinerant. In particular, it has no analog in FQH systems with continuous magnetic translation symmetries, and should be distinguished from ``anyon molecules" studied extensively recently~\cite{jain2025molecular,xu2025anyoncluster,khalaf2026bound,taige2026anyonmolecule}. Moreover, the weak-pairing $p-\mathrm{i}p$ CF superconductor is topologically nontrivial, and contributes $3$ chiral Majorana edge modes, since there are $3$ species of CFs. Therefore, in this case, we get
\begin{equation}
c_-=-2+\frac{3}{2}=-\frac{1}{2}.
\end{equation}

\section{ Parton Construction of the Low-Energy Effective Theory}
We provide a route to the effective theory Eqs.~\eqref{eq:CF_Lagrangian1},~\eqref{eq:CF_Lagrangian2} from a microscopic lattice model, keeping track of the electron operator so as to connect it to the fields in the final low-energy effective theory.  Consider a lattice model with hole filling of $2/3$ in each unit cell, which hosts an FCI phase.
It is helpful, although not necessary,  to think of the holes as being exposed to $2\pi$ flux within a unit cell, so that the vector potential they couple to is the combination: $A_e =A_{u.c.}+A$, where $\iint \nabla \times A_{u.c.} = 2\pi$ is responsible for the flux within the unit cell  and $A$ is the perturbation we probe the system with. 
We use a parton mean-field approach, decomposing the electron operator $\hat{c}=bf$ into a boson $b$ and fermion $f$ coupled to internal gauge fields $\alpha$ and $-\alpha+A_e$, respectively. 
Although the filling factor is fractional, by adjusting the flux of the internal gauge field $\frac{\nabla \times \alpha}{2\pi}=1/3$ per unit cell, the $f$ fermion density then matches the total flux they are exposed to: $\frac{\nabla \times (A_{u.c.}-\alpha)}{2\pi} =  2/3$, and can naturally form a Chern insulator with $\sigma_{xy}=1$, $c_-=1$ . At the same time, the $b$ bosons experience flux $1/3$ and are at density $2/3$ per unit cell, implying a filling of $\nu_b=2$ .  A natural insulating state at this filling is the bosonic integer quantum Hall state with Hall conductance $\sigma_{xy}=2$ and no net chirality~\cite{lu2012integer,senthil2013integer}. The parton construction aligns with the constructions in~\cite{taige2025chiral,zhang2026holon}, but differ from those in~\cite{shi2025doping,pichler2026microscopic,lotric2026}. 
Integrating out these partons leads to the low-energy theory for the undoped $2/3$ FCI:
\begin{equation}
\mathcal{L} = \frac{3}{4\pi}\alpha d\alpha -\frac1{2\pi}\alpha dA +\frac1{4\pi}AdA + 2\mathrm{CS}_g
\label{Eq:2/3} 
\end{equation}
The last two terms are background terms that add Hall conductance and a chiral central charge of unity. The $2\mathrm{CS}_g$ term  is referred to  as a ``gravitational Chern-Simons term", whose precise form will not be important to us, except that it is a useful bookkeeping device to keep track of the chiral central charge. Here $2\mathrm{CS}_g$  simply captures $c_-=1$. 

Then we discuss anyon excitations of the FCI. The excitation represented by $b$ partons carries statistics $\theta=-\frac\pi3$ and the minimal charge $q=e/3$, which can be verified from standard $K$ matrix techniques~\cite{wen1995topological}, so $b$ corresponds to the $1/3$ anyons. 
Because of the fractional emergent flux per unit cell, translations are implemented as magnetic translations on $b$, meaning that the elementary translations no longer commute but instead satisfy $T_1T_2=e^{-2\pi\mathrm{i}/3}T_2T_1$.  
It leads to three-fold degeneracy of the magnetic Bloch band of $b$.

Now consider the effect of finite doping. We will assume that charges enter in the form of the minimally charged anyons, i.e. the $b$ particles. Doping a density $\frac{\delta}{|\bm l_1\times \bm l_2|}$ of holes leads to $\frac{3\delta}{|\bm l_1\times \bm l_2|}$ $b$ particles. At $\delta\ll 1$, they enter the 3 degenerate valleys, and in the low-energy theory we get three fields $b_i$. Thus the effective Lagrangian at finite doping is
\begin{equation}
\mathcal{L}_b = \sum_{i=1}^3 b^\dagger_i D_\alpha b_i +\frac{3}{4\pi}\alpha d\alpha -\frac1{2\pi}\alpha dA +\frac1{4\pi}AdA + 2\mathrm{CS}_g. 
\end{equation}
If we choose the magnetic unit cell to be $\bm l_1\times 3\bm l_2$, then the lattice translations act as:
\begin{equation}
\begin{alignedat}{2}
T_1:~
b_i(\bm r)
&\rightarrow
e^{\mathrm{i}\bm K_i\cdot\bm l_1}
b_i(\bm r+\bm l_1),
&~
\alpha(\bm r)
&\rightarrow
\alpha(\bm r+\bm l_1),
\\
T_2:~
b_i(\bm r)
&\rightarrow
e^{\mathrm{i}\bm K_i\cdot\bm l_2}b_{i+1}(\bm r+\bm l_2),
&~
\alpha(\bm r)
&\rightarrow
\alpha(\bm r+\bm l_2).
\end{alignedat}
\label{eq:b_translation}
\end{equation}
The final step is to pass from the bosonic fields $b_i$ to CF fields $\psi_i$. 
The doped anyons see an effective magnetic field that puts each flavor at filling $\nu_{b_i}=-1$. Attaching one unit of flux to each $b_i$ converts them into fermions $\psi_i$ in zero field. We implement this via an additional parton construction $b_i=g_i\psi_i$ (Alternately, considering $b_i$ with the hard-core condition, a flux attachment transformation allows us to transform $b_i$ to CF $\psi_i$ exactly), introducing gauge fields $a'_i$ for each species, with $\alpha$ coupling to the $g_i$ fermions. Now $g_i$ are at unit filling $\nu_i=-1$, and can form a Chern insulator with $\sigma_{xy}=c_-=-1$; integrating them out leads to
\begin{equation}
\begin{aligned}
\mathcal{L}_\psi &=& \sum_{i=1}^3 \left [ \psi^\dagger_i D_{a'_i} \psi_i  -\frac1{4\pi}(-a'_i +\alpha) d(-a'_i+\alpha)\right ]\\ &&+\frac{3}{4\pi}\alpha d\alpha -\frac{1}{2\pi}\alpha dA +\frac{1}{4\pi}AdA -4\mathrm{CS}_g
\label{Eq:Final}
\end{aligned} 
\end{equation}
Encouragingly the self Chern-Simons term  for $\alpha$ cancels out and the field $\alpha$ now serves purely as a Lagrange multiplier. integrating it out forces the constraint $a'_3=-a'_1-a'_2+A$. Imposing this constraint leads us to the final theory:
\begin{equation}
\begin{aligned}
\mathcal{L}_\psi &=& \psi^\dagger_1 D_{a'_1} \psi_1 +\psi^\dagger_2 D_{a'_2} \psi_2+\psi^\dagger_3 D_{-a'_1-a'_2+A} \psi_3\\ &&- \frac{K_{IJ}}{4\pi}a'_I da'_J+\frac{a'_1+a'_2}{2\pi}dA -4\mathrm{CS}_g
\end{aligned} 
\label{Eq:Final}
\end{equation}
where $K$ is the same $2\times 2$ matrix as in Eq.~\eqref{eq:K_matrix}. The Lagrangian~\eqref{Eq:Final} was also derived in Ref.~\cite{zhang2026holon}.
Note this is equivalent to the form in Eq.~\eqref{eq:CF_Lagrangian2} but with $a'_I= a_I+A/3$, so that $a'_I$ are now properly quantized periodic gauge fields coupled to fermions. More precisely, these are spin$_c$ connections, whose flux quantization condition is fixed by the spin-charge relation, i.e. all charge $e$ local excitations must be fermions. 
The translation acts nontrivially both on $\psi_i$ and $a_I$:
\begin{equation}
\begin{alignedat}{2}
T_1:~
\psi_i(\bm r)
&\rightarrow
e^{\mathrm{i}\bm K_i\cdot\bm l_1}
\psi_i(\bm r+\bm l_1),
&~
a_I(\bm r)
&\rightarrow
a_I(\bm r+\bm l_1),
\\
T_2:~
\psi_i(\bm r)
&\rightarrow
e^{\mathrm{i}\bm K_i\cdot\bm l_2}
\psi_{i+1}(\bm r+\bm l_2),
&~
 a_I(\bm r)
&\rightarrow
W_{IJ}a_J(\bm r+\bm l_2).
\end{alignedat}
\label{eq:psi_translation}
\end{equation}
where 
$$W=\begin{pmatrix} 0 & 1\\ -1 & -1\end{pmatrix}$$
satisfies $W^TKW=K$, so the Lagrangian $\mathcal{L}_\psi$ is compatible with translation symmetry. 
In Appendix~\ref{app:generalizations} we show how to extend the above the construction to fillings $\nu=\frac{2n}{2n+1}$, and also discuss other possible phases of the Lagrangian \eqref{Eq:Final}.

\section{Discussion} 
Here we discuss the implications of our theory for recent experiments and numerical studies.
A superconducting phase has been observed recently adjacent to the $2/3$-filling FCI state in twisted MoTe$_2$~\cite{xu2025signatures}. Its proximity to an FCI makes the anyon-superconductor interpretation appealing. In Appendix~\ref{app:normal_state}, we analyze the transport data at temperatures above $T_c$ but well below the FCI gap. Our analysis suggests that the normal-state charge carriers are charged excitations of the FCI state, which supports the identification of the normal state as an itinerant anyon gas. Nevertheless, further experiments are necessary to determine whether this is really the case. Once an anyon superconductor scenario is adopted, a central question concerns how $1/3$ anyons pair into $2/3$ anyons: Previous theories take this as an assumption or find anyon molecules in the Landau-level setting~\cite{jain2025molecular,xu2025anyoncluster,khalaf2026bound,taige2026anyonmolecule}. Here we find that the pairing directly arises from the anyon kinetic energy and anyon statistics, without the need for additional pairing interactions.  

In our theory, the only energy scale is $\frac{2\pi \hbar^2 \bar n}{m}$, which sets the degeneracy temperature of the anyon gas. Our theory therefore predicts that, on the FCI side of the superconducting dome, both the zero-temperature superfluid stiffness and the superconducting gap are roughly proportional to doping $\delta = \nu - 2/3$, in the limit of small $|\delta|$. In reality, because of lattice effects, disorder, and interactions between anyons, which are neglected in our theory, the doping dependence will not be strictly linear. Indeed, experimentally, superconductivity sets in only above a finite doping~\cite{xu2025signatures}. Still, if pairing is mainly driven by the kinetic energy of $1/3$ anyons, we expect both the superfluid stiffness and the superconducting gap to be of the same order as $k_BT_c$: they should be small near the onset of the dome and increase with doping in the underdoped region. Other mechanisms lead to different predictions. For instance, in a superconductor formed by $2/3$ anyons~\cite{shi2025doping,nosov2026anyonsc}, $T_c$ in this region is limited only by the phase stiffness, and the superconducting gap should be much larger than $k_BT_c$. On the other hand, in a Kohn--Luttinger type superconductor~\cite{zhang2025chiralSC,guerci2025chiralSC}, the superfluid stiffness should be much larger than $k_BT_c$.
The doping dependence of the superfluid stiffness therefore probes whether the charge carriers are anyons doped into the FCI, while that of the superconducting gap can provide evidence for the kinetic-energy-driven pairing expected in our proposed mechanism of anyon superconductivity.

Recent numerics also show evidence of chiral superconductivity near the $\nu =2/3$ FCI~\cite{taige2025chiral,guerci2026topological}. Its chiral central charge is found to be $c_-=-1/2$, in disagreement with the previous theoretical prediction $c_-=-2$. Some works invoked more elaborate non-Abelian parton constructions to accommodate half-integer chiral central charge~\cite{Ma_2020, shi2025nonabelian,lotric2026}, but still fail to obtain the precise value $c_-=-1/2$. Our theory resolves this discrepancy. 
Our flux-attachment construction provides a unified framework encompassing both earlier approaches and ours: the earlier theories realize a strong-pairing regime, whereas our theory describes a weak-pairing regime, which is arguably the more natural outcome for itinerant anyons with kinetic energy.
The weak pairing phase yields $c_- = -1/2$ and physical pairing symmetry $f-\mathrm{i}f$, both in agreement with recent numerics~\cite{guerci2026topological,taige2025chiral}. Since $c_-$ is directly related to the thermal Hall conductance, we expect future thermal Hall measurements to provide a direct experimental test of our theory.

Our theory can be readily generalized to doped $\nu =1-1/m$ FCIs, with odd integer $m$. The doped anyons have statistics $\theta =- \pi/m$, charge $e/m$, and $m$ degenerate pockets. Again, we introduce a CF $\psi_i$ for the $i^{th}$ pocket, and choose the $m\times m$ flux-attachment matrix to be $\Phi_{ij}= 2\pi (\delta_{ij}-1/m)$. It can be implemented by a $U(1)^{m-1}$ Chern-Simons theory~\eqref{eq:CF_Lagrangian1}, with $I=1,2,\cdots,m-1$ (the rank of $\Phi_{ij}$ is $m-1$, so we need $m-1$ gauge fields), and 
\begin{equation*}
K_{IJ}=(\delta_{IJ}+1),    \qquad     q_{iI}=\delta_{iI}-\delta_{im}\,.
\end{equation*}
Following the same analysis as in the $m=3$ case (the main focus of this work), we can show that CFs can enter an inter-pocket $p-\mathrm{i}p$ paired state, leading to physical superconductivity. The resulting chiral central charge is $c_- = 1-m/2$ in the weak pairing phase, while in the strong pairing phase we obtain $c_-=1-m$. The corresponding values for filling near $\nu= 1/m$ can be obtained by a particle hole transformation, which sends $c_-\rightarrow 1-c_-$. This construction suggests that doping FCIs at fillings $\nu = 1/m$ and $\nu = 1-1/m$ provides a broad and promising route to chiral superconductivity. Testing this possibility in future experiments on FCIs beyond $2/3$ filling would be an important direction.

\section{Acknowledgements}
 We thank Jing-Yuan Chen, Tingxin Li and Xiaoxue Liu for helpful discussions. Z.W. is especially grateful to Fan Xu for sending the experimental data of Ref.~\cite{xu2025signatures}. A.V. acknowledges insightful discussions and prior collaborations with Max Metlitski and Clemens Kuhlenkamp. 
This work was supported in part by the National Natural Science Foundation of China, Grant No. 12125405 (Z.W.), and by the Simons Collaboration
on Ultra-Quantum Matter, which is a grant from the Simons Foundation (651440, A.V., Z.F.).

\bibliographystyle{apsrev4-2}
\bibliography{ref}

\newpage
\appendix

\section{Flux attachment in first quantization\label{app:flux_attachment}}
In this section, we provide a derivation of the flux-attachment transformation in the first-quantization language, including the subtlety about inter-pocket statistics. Our derivation follows~\cite{zhang1992cslg2,laughlin1988sc}.
In the following, we use a phenomenological description of anyon gas emerging from the doped FCI. We start with the Hilbert space of $N\in 3\mathbb{Z}$ anyons, and treat anyon pocket as a flavor index $i=1,2,3$. We assume $N_1=N_2=N_3=N/3$. 
We define the basis state $|I;X\rangle$, where $X=(\mathbf{r}_1,\mathbf{r}_2,\cdots \mathbf{r}_N)$ denotes the collection of anyon coordinates, and $I=i_1i_2\cdots i_N$ is the collection of the pocket indices. Anyon statistics  $\theta =-\pi/3$ requires that
\begin{equation}
|I;X\rangle \to e^{-\mathrm{i}\pi/3}|I;X\rangle
~\text{under}~\mathbf r_a \,\ccwexchange\, \mathbf r_b,
~ i_a\leftrightarrow i_b
\label{eq:exchange_basis_state}
\end{equation}
We denote the first-quantized anyon wavefunction as $\Psi^{\rm{anyon}}_I(X)$, and impose the hard-core constraint 
\begin{equation}
\Psi_I^{\rm anyon} (\cdots, \bm r_a=\bm r_b,\cdots)=0 .
\label{eq:hardcore}
\end{equation}
From Eq.~\eqref{eq:exchange_basis_state}, we have
\begin{equation}
\Psi^{\rm{anyon}}_I(X)\to e^{\mathrm{i}\pi/3}\Psi^{\rm{anyon}}_I(X)
~\text{under}~ \mathbf r_a \,\ccwexchange\, \mathbf r_b,
~ i_a\leftrightarrow i_b
\label{eq:exchange_wavefunction}
\end{equation}
In particular, the wavefunction $\Psi^{\rm{anyon}}_I(X)$ is not single-valued due to nontrivial braiding statistics. In this basis, the Hamiltonian of the non-interacting anyon gas is simply given by:
\[
\hat H_{\rm{anyon}}=\sum_{a=1}^{N} \frac{\hat p^2_a}{2m}\,.
\]

We define following non-local transformation 
\begin{equation}
\Psi^{\rm{anyon}}_I(X)=\eta_I U(I,X)\Psi^{\rm{CF}}_I(X)
\end{equation}
to transform $\Psi$ into the composite-fermion wavefunction $\Psi^{\rm{CF}}$, where 
\begin{equation}
\begin{gathered}
U(I,X)=\prod_{a<b}\exp \big(-\mathrm{i}\frac{\Phi_{i_ai_b}}{2\pi} \phi_{\bm r_a-\bm r_b}\big)\, \\
\left(\frac{\Phi_{ij}}{2\pi}\equiv\delta_{ij}-\frac{1}{3},~\phi_{\bm r} \equiv\text{ polar angle of } \bm r\right)
\end{gathered}
\end{equation}
is the flux attachment phase factor. As mentioned in the main text, the flux attachment renders the braiding statistics of CFs trivial and intra-pocket exchange statistics fermionic. To fulfill the inter-pocket fermionic statistics, i.e.,
\begin{equation}
\Psi^{\rm{CF}}_I(X)\to -\Psi^{\rm{CF}}_I(X)
\quad \text{under}\quad \mathbf r_a \,\ccwexchange\, \mathbf r_b,
\quad i_a\leftrightarrow i_b\,,
\end{equation}
we need to introduce an additional sign factor $\eta_I$, defined as 
\begin{equation}
\eta_{I}=\begin{cases} +1 \text{ if } I \to I_0 \text{ takes even permutations }\\ -1 \text{ if } I \to I_0\text{ takes odd permutations }\end{cases} 
\end{equation}
where $I_0\equiv 11\cdots 1122\cdots 2233\cdots33 $. Since the Hamiltonian is diagonal in the pocket index $\langle I;X'|H|J\neq I;X\rangle =0$, this sign factor does not appear in the CF Hamiltonian, consistent with the expectation that only full braiding statistics are physically meaningful for distinguishable particles. Then we obtain 
\begin{equation}
\begin{gathered}
\hat H_{\rm{CF}}=\hat U^\dagger \hat H_{\text{anyon}} \hat U=\frac{1}{2m}\left(\hat p_b-\hat{\bm{a}}_b\right)^2,\\
\hat{\bm{a}}_b =\sum_c \frac{\Phi_{i_bi_c}}{2\pi} \cdot\frac{\hat z\times(\hat{\bm r}_b-\hat{\bm r}_c)}{|\hat{\bm r}_b-\hat{\bm r}_c|^2} \,.
\end{gathered}
\end{equation}
The second quantized version of $\hat H_{\rm{CF}}$ is exactly Eq.~\eqref{eq:H_CF} in the main text. 

Finally, as is pointed out in Ref.~\cite{Kim1997contact}, the hard-core condition~\eqref{eq:hardcore} can also be implemented dynamically, by adding a contact interaction term
\begin{equation}
\frac{2\pi}{3m}\sum_{a<b}\delta^{(2)}(\hat {\bm r}_a-\hat{\bm r}_b)
\end{equation}
to the Hamiltonian (both for $\hat H_{\rm CF} $ and $\hat H_{\rm anyon}$), which makes the connection to the field theory description clear.

\section{Particle-Vortex Duality\label{app:pv_duality}}

Here we use particle-vortex duality to explicitly show the absence of deconfined anyons in the inter-pocket paired state~\cite{peskin1978duality,halperin1981duality}.  We start from the phase-only action in Eq.~\eqref{eq:CSLG}. For convenience, we set the velocity of the Goldstone mode $v=1$, and use the short-hand $X_\mu ^2\equiv -X_0^2+\bm X^2$.
We first perform a Hubbard-Stratonovich transformation:
\begin{equation}
-\frac{\chi_s}{2}(\partial_\mu \theta_p-\alpha_{p\mu } )^2
\rightarrow \frac{1}{2\chi_s}(J_{p\mu})^2  +J^\mu_p(\partial_\mu \theta_p-\alpha_{p\mu}) \,.
\label{eq:particle_vortex_1}
\end{equation}
Then we separate the phase field into a smooth part and a vortex part: $\theta_p =\theta^{s}_p+\theta^v_p$. Integrating out $\theta^s_p$, we obtain the constraint $\partial_\mu J^\mu_p =0$, which can be solved by introducing a dual gauge field $b_p$ for each
pair field, $J_p^\mu=\frac{1}{2\pi}\epsilon^{\mu\nu\lambda}\partial_\nu b_{p\lambda}$, and the source term of $b_p$ is the vortex current $j^{v\mu}_p = \frac{1}{2\pi}\epsilon^{\mu\nu\lambda} \partial_\nu\partial_\lambda
\theta^v_p$.
Dropping the less relevant Maxwell term $\propto (db_p)^2$, we get the particle-vortex dual of Eq.~\eqref{eq:CSLG}
\begin{equation}
\begin{gathered}
{\cal L}_{\text{dual}}= \sum_p-{1\over 2\pi}b_p d \alpha_p +b_{p\mu} j^{v\mu}_p-{1\over 4\pi}K_{IJ}a_Ida_J+\frac{2/3}{4\pi }AdA
\end{gathered}
\end{equation}
It is useful to separate the two relative phase modes from the total phase mode by the unimodular change of variables
\begin{equation}
\beta_1=b_{12}-b_{23},\qquad
\beta_2=b_{12}-b_{31},\qquad
\beta_3=b_{12}.
\end{equation}
Then
\begin{equation}
\begin{aligned}
{\cal L}_{\text{dual}}=&
-{1\over 2\pi}\left(\beta_1da'_1+\beta_2da'_2\right)
-{1\over 4\pi}K_{IJ}a'_Ida'_J\\
&+\frac{1}{2\pi}(a'_1+  a'_2+\beta_1+\beta_2-2\beta_3)dA\\
&-\beta_{1\mu} j^{v\mu }_{23}-\beta_{2\mu} j^{v\mu }_{31}+\beta_{3\mu}j^{v\mu }_\Theta\,,
\end{aligned}
\end{equation}
where $a'_I\equiv a_I+A/3$. The dynamical topological sector involving $(a'_1,a'_2,\beta_1,\beta_2)$ is described by
$
{\cal K}=
\begin{pmatrix}
K&I\\
I&0
\end{pmatrix}.
$
Since $\det\mathcal{K}=1$, each vortex excitation of the relative phase mode binds an integer number of gauge flux quanta, and therefore has trivial braiding statistics. 

The remaining dual field $\beta_3$ is only coupled to $A$ and the vortex current of the global $U(1)$ phase $\Theta$, $j^{v\mu}_\Theta=\frac{1}{2\pi}\epsilon^{\mu\nu\lambda} \partial_\nu\partial_\lambda
\Theta$. The equation of motion of $\beta_{3}$ gives:
\begin{equation}
j^v_\Theta={2\over 2\pi}dA .
\end{equation}
Thus a unit vortex carries magnetic flux $h/2e$.  This is consistent with the flux-quantization argument in the main text.

\section{Normal-state transport in twisted MoTe$_2$\label{app:normal_state}}

In~\cite{xu2025signatures}, a superconducting phase was observed in twisted MoTe$_2$ at twist angle
$3.83^\circ$, in the hole-filling region $0.71\leq \nu_h \leq 0.76$, next to the $\nu_h =2/3$ FCI state. Our theory offers a natural explanation for this superconductivity. To understand the nature of the superconducting state, one
needs to determine whether the normal state at $T_c<T\ll \Delta_{\text{FCI}}  ~(\text{gap of the FCI})$ realizes an itinerant anyon gas. Such an anyon gas is expected slightly away from the Hall plateau. A seemingly puzzling feature of the experiment is that the normal-state longitudinal resistivity $\rho_{xx}$ exhibits a peak between $\nu_h=2/3$ and $\nu_h\simeq 0.71$, which might appear to argue against an anyon-gas interpretation.

\begin{figure}[htb]
    \centering
    \includegraphics[width=0.8\linewidth]{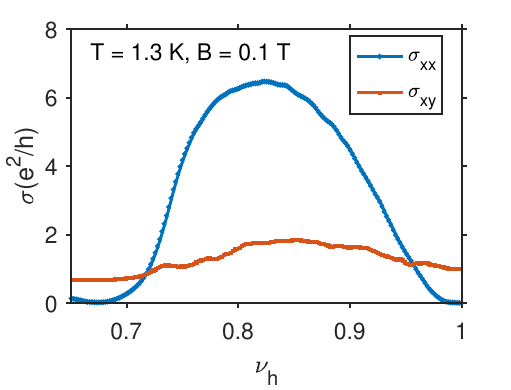}
    \caption{
    Conductivities $\sigma_{xx}$ and $\sigma_{xy}$ obtained from the
    normal-state resistivity data in Ref.~\cite{xu2025signatures} at
    $T=1.3\,\mathrm{K}$ and $B=0.1\,\mathrm{T}$.
    Since the original $\rho_{xx}$ and $\rho_{xy}$ data are given at
    slightly different filling factors, we linearly interpolate $\rho_{xy}$
    onto the original filling-factor grid of $\rho_{xx}$ within their common
    filling range before performing the tensor inversion. 
    }
    \label{fig:conductivity}
\end{figure}

However, this feature is much less mysterious when expressed in terms of the conductivity, by inverting the resistivity tensor,
\begin{equation}
\sigma_{xx}
=
\frac{\rho_{xx}}{\rho_{xx}^2+\rho_{xy}^2},
\qquad
\sigma_{xy}
=
-\frac{\rho_{xy}}{\rho_{xx}^2+\rho_{xy}^2}.
\end{equation}
In Fig.~\ref{fig:conductivity}, we plot the resulting conductivities as functions of the hole filling factor $\nu_h$, based on the experimental normal-state resistivity data at $B=0.1\,\mathrm{T}$. We choose $T=1.3\,\mathrm{K}$, which is slightly above the highest onset transition temperature. At $\nu =2/3$, $\sigma_{xx}$ is negligible, indicating that $T\ll \Delta_{\text{FCI}}$. At $\nu_h\lesssim 0.76$, the longitudinal conductivity increases approximately linearly with doping once away from the Hall plateau. The peak in $\rho_{xx}$ between filling $2/3$ and $0.71$ is a simple consequence of the monotonic growth of $\sigma_{xx}$, given the relation $\rho_{xx}=\frac{\sigma_{xx}}{\sigma^2_{xx}+\sigma^2_{xy}}$. This behavior suggests that the mobile charge carriers are charged excitations of the parent FCI. Therefore, the normal state is likely to be an anyon gas, at least on the low-doping side of the superconducting dome. Nevertheless, further experiments are necessary to determine whether the charge carriers are indeed anyons.

Notably, the behavior of longitudinal conductivity is consistent with Eq.~\eqref{eq:phys_conductivity} applied to the normal state, which is mapped to a CF metal under our flux-attachment construction. In reality, the anyon band can have nonzero Berry curvature, which has been neglected in our analysis. Taking this into account, the CFs should form an anomalous Hall metal at $T>T_c$. Then according to Eq.~\eqref{eq:phys_conductivity} in the main text, nontrivial anomalous Hall transport of CFs can also lead to an increase of $\sigma_{xy}$ with doping.

\section{Generalizing to the Laughlin sequence: the parton field theory approach\label{app:generalizations}}
 We start again with $\hat{c} = bf$, now at filling $2n/(2n+1)$ (equivalent to the Laughlin sequence $1/(2n+1)$ by a particle hole transformation). The  gauge field $\alpha$ couples to $b,\,f$ with charge $+1,\,-1$. The electric charge of $A$ is carried by $f$. Again we introduce a background flux $\iint \nabla \times \alpha = 2\pi/(2n+1)$ in each unit cell, which has two effects. It places the $f$ fermions in an integer quantum Hall state with $C=1$ and the bosons are at filling $2n$ where they can form a bosonic integer quantum Hall phase with $\sigma_{xy}=2n$ and no chiral edge modes. The Ioffe-Larkin rule of combining conductivities of partons tells us the Hall conductance of the electrons for this state is $\sigma_{xy}^{-1} = 1+\frac1{2n}=\frac{2n+1}{2n}$ i.e. the required Hall conductance. Also, we have $2n+1$ pockets for the $b$ bosons on doping, because they couple minimally to the $\alpha$ which have fractional flux per unit cell.

This gives us the effective theory:
\begin{equation}
\mathcal{L}_b = \sum_{i=1}^{2n+1} b^\dagger_i D_\alpha b_i +\frac{2n+1}{4\pi}\alpha \,d\alpha  -\frac1{4\pi}\alpha \,dA +\frac1{4\pi}A\,dA + 2\text{CS}_g
\label{Eq:2/3} 
\end{equation}

Again we find that each boson species is at filling $\nu_i=-1$. We therefore perform a flux attachment to convert the bosons into composite fermions through the successive parton  construction: $b_i=g_i\psi_i$ and in the process introduce $a_i$ with $i=1,\dots 2n+1$. We choose that the gauge charge of $\alpha$ is carried by the $g_i$. The $g_i$ are put in $C=-1$ Chern insulators and can be integrated out giving:

\begin{eqnarray}
\mathcal{L}_\psi &=& \sum_{i} \left [ \psi^\dagger_i D_{a_i} \psi_i  -\frac1{4\pi}(-a_i +\alpha) d(-a_i+\alpha)\right ]\\ 
&&+\frac{2n+1}{4\pi}\alpha d\alpha -\frac1{4\pi}\alpha dA +\frac{A}{4\pi}dA -2(2n) \text{CS}_g \nonumber
\label{Eq:Final2}
\end{eqnarray}

Importantly, the term $\alpha d\alpha$ cancels out, so the field $\alpha$ appears only in linear order and we can integrate out, which enforces the constraint $a_{2n+1} = A-\sum_{i=1}^{2n} a_i $. Putting this all together we get the following (where we defined the restricted sum: $\sum'_i = \sum_{i=1}^{2n}$) :

\begin{eqnarray}
\mathcal{L}_\psi &=&\sum_i{'} \psi^\dagger_i D_{a_i} \psi_i + \psi^\dagger_{2n+1} D_{A-\sum_i' a_i} \psi_{2n+1} \\ &&- \frac{K_{IJ}}{4\pi}a_I da_J+\frac{\sum_i' a_i}{2\pi}dA -4n \text{CS}_g \nonumber
\label{Eq:Final3}
\end{eqnarray}
where:
 $K 
=\begin{pmatrix}
2 & 1 & 1 & \cdots & 1\\
1 & 2 & 1 & \cdots & 1\\
1 & 1 & 2 & \cdots & 1\\
\vdots & \vdots & \vdots & \ddots & \vdots\\
1 & 1 & 1 & \cdots & 2
\end{pmatrix} \equiv I_{2n}+J_{2n}
$
Where $I_{2n}$ is the $2n\times 2n$ identity matrix and $J_{2n}$ is the all ones matrix, 
$(J_{2n})_{ij}=1\ \ \forall\, i,j\in\{1,\dots,2n\}$. This matrix is readily diagonalized since $J$ is a rank-one matrix with eigenvalues $(2n,\,0,\dots 0)$ and hence ${\rm Det} K =2n+1$, and ${\rm Signature} (K)  = 2n$. This gives us the topological properties of the $\nu=2n/(2n+1)$, the particle hole conjugate of the Laughlin state with $m=2n+1$, including vanishing chiral central charge, the correct statistics of anyons and their charges.

Finally, we pair between pockets and generalize the $M_{ij}$ matrix.
We assume that $\operatorname{rank}M=2n+1$ so that the Fermi surfaces of $\psi_i$'s are fully gapped. We further assume adjacent pockets have nonvanishing pairing $\Delta_{i,\,i+1}$, with $i+1$ defined modulo $2n+1$ (due to the permutation symmetry among valleys, there is at least one ground state with nonvanishing $\Delta_{i,i+1}$). This leads to the fully Higgsed scenario and the order parameter 
$$\mathcal{O}_{\mathrm{SC}}(\bm R,\bm r)=e^{2i\sum_i \bm K_i\cdot \bm R} \prod_{i=1}^{2n+1}\Delta_{i,i+1}(\bm R,\bm r)
\,,$$
which is gauge invariant with respect to the internal gauge fields and carries charge $2e$, i.e. a paired superconductor with {\em no} residual topological order. Now, if we again pick $p-ip$ weak pairing between pockets, we will get chiral central charge of $\Delta c_-= (2n+1)/2$ so an additional term in the action  which leads to the combined form: $ (2n+1)\text{CS}_g -4n\text{CS}_g=(1-2n) \text{CS}_g$. Thus the chiral central charge of the weak pairing superconductor is $c_- = -(2n-1)/2$ when the charge Hall conductance is $\sigma_{xy} = \frac{2n}{2n+1}$. For $n=1$ this reduces to the usual $c_-=-1/2$. The strong pairing of composite fermions leads to chiral central charge at the edge of $c_-=-2n$. 

\section{Charge Density Wave States\label{app:CDW}}
In this section, we explore other possible phases exposed under our CF description of the 3-pocket $1/3$ anyon gas. 
Consider now states which break the valley permutation symmetry - for instance that pick a pair of valleys and induce pairing just between them. This choice effectively breaks translation symmetry generated by $T_1$ (which permutes the three valleys), and corresponds to period 3 density waves. For instance, say we pair the composite fermions $\psi_i$ only between $i=1,\,2$ and leave $\psi_3$ unpaired. This assumption can be made without loss of generality given the permutation symmetry of the valleys. Further, the pairing between these pockets can have a winding of order $k$ which leads to topological responses in the weak pairing limit. Denoting $\langle \psi _1(-\bm k) \psi_2(\bm k) \rangle =|\Delta_{12}(\bm k)|e^{i\theta_{12}}$. This pairing has two effects - first it Higgses the gauge potentials due to the condensate $-\frac{\chi}{2}(\partial_\mu \theta_{12} -a'_{1\mu}-a'_{2\mu})^2 $ which leads to the Meissner effect locking together $a'_1$ and $a'_2$, so we can define $a'_1=-a'_2=a$. Second, it leads to a topological term $\frac{k}{4\pi}ada +2k \mathrm{CS}_g$, which follows from performing a particle-hole conjugation of $\psi_2$ following which both species of fermions couple with the same charge to $a$.  This converts the topological superconductor into a Chern insulator with Chern number $k$ and $c_-=k$. This is precisely the term written above. Putting this together we have for the $2$-pocket pairing theory:
\begin{equation}
\mathcal{L}_{12} =\frac{k-2}{4\pi} ada+(2k-4)\mathrm{CS}_g+\psi_3^\dagger D_A \psi_3 
 \end{equation}

Based on the previous energetics arguments, we restrict attention to $k=0,\,1$ although for completeness we also consider $k=2$. 

 (i) For $k=0$ corresponding to $s$-wave or strong pairing between the two pockets, we see that there is residual ``semion" topological order $\frac2{4\pi}ada$, coexisting with the CDW and doped holes. This is nothing but the CDW state with ``dark" topological order dubbed the ``semion crystal"  \cite{Song_2024,pichler2026microscopic,zhang2026holon} in recent works. 
 (ii) For $k=1$ ($p-\mathrm{i}p$, weak pairing), the topological term $-\frac1{4\pi}ada-2\mathrm{CS}_g$ is known to be trivial and can be integrated out \cite{seiberg2016}, leaving behind a CDW without  topological order and a metallic pocket of doped holes. Interestingly this state has no background Hall conductance, which is challenging to access from other parton approaches. 
 (iii) For $k=2$ we see that the topological term and chiral central charge are entirely canceled, again pointing to the trivial CDW metal. Interestingly, this state can be seen to break both $T_1$ and $T_2$. Note that $T_1:\psi_i\rightarrow e^{2\pi \mathrm{i} s_i/3}\psi_i$, where $s_i=3\bm K_i\cdot \bm l_1/2\pi $, and $s_{i+1}=s_i+1~(\text{mod}~3)$. We introduce a background field $B$ for translation generated by $T_1$, so that $\psi_i$ is coupled to $s_iB$. The condensate $\Delta_{12}$ now imposes $a'_1+s_1B=-a'_2-s_2B\equiv a$. Then integrating out $\psi_1,\psi_2$ leads to an additional response $-\frac{2}{4\pi}adB$. For $k=2$, $a$ becomes a Lagrangian multiplier, and enforces $dB=0$, i.e., dislocations have divergent energy cost. This indicates symmetry breaking of $T_1$, in addition to the apparent $T_2$ symmetry breaking due to two-pocket pairing.

\bigskip

\end{document}